\documentstyle[12pt]{article}
\begin{document}
\title{
 A Parity Invariant Regularization \\
 in\\
 3-D Quantum Electrodynamics
}
\author{{\sc Tadahiko Kimura}%
  \thanks{E-mail address: kimura@cuphd.nd.chiba-u.ac.jp}
  \vspace*{0.5em}
  \\
  {\sl Department of Physics, Faculty of Science, Chiba University }\\
  {\sl 1-33 Yayoi-cho, Inage-ku, Chiba 263, Japan} \\
}
\date{
 CHIBA-EP-76 \\
 hep-th/9404024 \\
 April 1994
}
\maketitle
\begin{abstract}
We regularize the 3-D quantum electrodynamics by a
parity invariant Pauli-Villars regularization method. We find that in
the perturbation theory the Chern-Simons term is not induced in the
massless fermion case and induced in the massive fermion case.
\end{abstract}

\newpage
There has been much recent interest in Chern-Simons theories \cite{JT}
in three dimensions in several contexts; their ability \cite{LFC} to
explain certain behaviour in fractional quantum Hall effect and high
$T_{c}$ superconductivity, their connection \cite{WP} with knots and
conformal field theories in 2-D, snd so on.

One of the important problems when studying Chern-Simons theories is
the method employed to regularize ultraviolet divergent integrals.
Three attempts have been made to regularize the theories; i) the
Pauli-Villars regularization \cite{DJT} \cite{ALR}, ii) the analytic
regularization \cite{PST} and iii) the dimensional regularization
\cite{M} \cite{DW}. Unfortunately the results have not been totally
conclusive, and all or some of these regularization metdods seem to
suffer from their own difficulties. One important such discrepancy which
 is concerned in this note is whether the Chern-Simons terms are induced
through quantum fermion loop corrections even if the bare Lagrangian
does not contain such terms. Untill now it has been argued \cite{DJT}
\cite{PST} \cite{DW} that for massless fermions Chern-Simons term is
induced in the  Pauli-Villars regularization and not induced
in the analytic and dimensional ones, and for massive fermions the
Chern-Simons term is not induced in the Pauli-Villars regularization and
 induced in the analytic and dimensional ones. The reason why the
Chern-Simons term is (is not) induced for massless (massive) fermions in
 the Pauli-Villars regularization method is the following. The usual
Pauli-Villars regularization in 3-D breaks parity due to the regulator
fermion mass and the massive regulator fermion loop induces the
Chern-Simons term which remains finite even after the regulator masses
tend to infinity. For massive fermion case this induced Chern-Simons
term cancels the one induced  through original fermion loop, and for
massless fermion case it survives. Therefore it is desirable to develop
a parity invariant Pauli-Villars regularization method in which the
regulator fields do not induce any Chern-Simmons term to discuss whether
 the Chern-Simons terms are induced through the quantum fermion loop
correction or not.

On the other hand recently Frolov and Slavnov \cite{FS} has proposed
a method of chiral gauge invariant regularization for the fermion field
in the 16-component irreducible spinor representation of SO(10) chiral
gauge theory, which is automatically anomaly-free. The method is based
on the introduction of infinite number of the Pauli-Villars regulator
spinor fields which is in the 32-component reducible representation of
SO(10). The method seems to work for any anomaly-free chiral gauge
theories \cite{NN}.

\vspace{5mm}
The purpose of this short note is to clarify this problem in the
perturbation theory by adapting a parity invariant Pauli-Villars
regularization method which is a kind of 3-D variant of the chiral gauge
invariant Pauli-Villars regularization proposed by Frolov and Slavnov
\cite{FS}. Our conclusion is that this parity invariant Pauli-Villars
regularization gives the same result as that in analytical and
dimensional regularization methods.

\vspace{5mm}
We consider the 3-D quantum electrodynamics with one flavour of
two-component spinor $\psi$ of mass $m$. Note that one two-component
spinor is counted as one flavour. In order to regularize the ultraviolet
 divergenct integrals in parity invariant way, following Frolov and
Slavnov, we must introduce the infinite number of fermionic and bosonic
Pauli-Villars regulator spinor fields which belong to the 4-component
reducible representation of the 3-D Lorentz group; $\psi_{r}$ and $\phi_
{s}$, respectively. The fermions with the odd number of flavours cannot be
regularized by the regulator fields with the finite even number of
flavours. In the case of even number of flavours it is sufficient to
introduce the finite number (a half number of flavours) of 4-component
regulator spinor fields. With the two-component irreducible spinors they
are written as $\psi_{r}=(\psi_{+r}, \psi_{-r})$ and $\phi_{s}=(\phi_
{+s}, \phi_{-s})$. The parity transformation of the two-component
spinors $\psi$, $\psi_{r\pm}$ and $\phi_{s\pm}$ is
defined as

\begin{eqnarray}
x=(x_{0},x_{1},x_{2})\rightarrow \bar{x}=(x_{0},-x_{1},x_{2}),
\nonumber \\
 \psi (x) \rightarrow \sigma_{1}\psi (\bar{x}), \psi_{r\pm}(x)
\rightarrow \sigma_{1}\psi_{r\mp}(\bar{x}), \phi_{s\pm}(x) \rightarrow
\sigma_{1}\phi_{s\mp}(\bar{x}).
\end{eqnarray}
Then the Lagarangian density regularized in a gauge and parity invariant
 way is given by

\begin{eqnarray}
 L&=& - \frac{1}{4}F_{\mu\nu}^{2} + \bar{\psi}i{\gamma}^{\mu}D_{\mu}
\psi + m\bar{\psi}\psi  \nonumber  \\
  & & +\sum_{r=1}^{\infty}\{ \bar{\psi}_{r+}i\gamma^{\mu}D_{\mu}\psi_
{r+} + \bar{\psi}_{r-}i\gamma^{\mu}D_{\mu}\psi_{r-} + m_{r}(\bar{\psi}
_{r+}\psi_{r+}- \bar{\psi}_{r-}\psi_{r-})\}  \nonumber \\
  & & +\sum_{s=1}^{\infty}\{ \bar{\phi}_{s+}i\gamma^{\mu}D_{\mu}\phi_
{s+} + \bar{\phi}_{s-}i\gamma^{\mu}D_{\mu}\phi_{s-} + m_{s}(\bar{\phi}
_{s+}\phi_{s+}- \bar{\phi}_{s-}\phi_{s-})\},
\end{eqnarray}
where $D_{\mu}\equiv\partial_{\mu}+ieA_{\mu}$ , $\bar{\psi}\equiv
\psi^{\dag}\sigma_{3}$, and $m_{r}$ and $m_{s}$ are masses of regulator
fields. Of course the mass term of original two-component spinor $\psi$
breaks the parity invariance. Our notaions used in this note are the
followings;

\begin{eqnarray}
\gamma_{\mu}\gamma_{\nu}+\gamma_{\nu}\gamma_{\mu}=2g_{\mu\nu},
diag[g_{\mu\nu}] = (+1,-1,-1) , \nonumber \\
\gamma_{0}=\sigma_{3}, \gamma_{1}=i\sigma_{1}, \gamma_{2}=i\sigma_{2},
\\
Tr[\gamma_{\mu}\gamma_{\nu}\gamma_{\rho}\gamma_{\sigma}]=2(g_{\mu\nu}
g_{\rho\sigma}+g_{\mu\sigma}g_{\nu\rho}-g_{\mu\rho}g_{\nu\sigma}),
\nonumber \\
Tr[\gamma_{\mu}\gamma_{\nu}\gamma_{\rho}]=2i\varepsilon_{\mu\nu\rho},
\varepsilon_{012}=-1,
\end{eqnarray}
where $\sigma's$ are Pauli matrices.

Now let us calculate the one loop quantum correction to the regularized
photon vacuum polarization tensor $\Pi^{M}_{\mu\nu}(p)$.

\begin{eqnarray}
\Pi^{M}_{\mu\nu}(p)&=& -e^{2}\int \frac{d^{3}k}{(2\pi)^{3}}\{Tr[\gamma_
{\mu}\frac{\gamma^{\rho}k_{\rho}-m}{k^{2}-m^{2}}\gamma_{\nu}\frac
{\gamma^{\sigma}(k+p)_{\sigma}-m}{(k+p)^{2}-m^{2}}] \nonumber \\
& & +\sum_{r}Tr[\gamma_{\mu}\frac{\gamma^{\rho}k_{\rho}-m_{r}}{k^{2}-m^
{2}_{r}}\gamma_{\nu}\frac{\gamma^{\sigma}(k+p)_{\sigma}-m_{r}}
{(k+p)^{2}-m^{2}_{r}} \nonumber \\
& & +\gamma_{\mu}\frac{\gamma^{\rho}k_{\rho}+m_{r}}{k^{2}-m^{2}_{r}}
\gamma_{\nu}\frac{\gamma^{\sigma}(k+p)_{\sigma}+m_{r}}{(k+p)^{2}-m^{2}_
{r}}] \nonumber \\
& & -\sum_{s}Tr[\gamma_{\mu}\frac{\gamma^{\rho}k_{\rho}-m_{s}}{k^{2}-m^
{2}_{s}}\gamma_{\nu}\frac{\gamma^{\sigma}(k+p)_{\sigma}-m_{s}}{(k+p)^{2}
-m^{2}_{s}} \nonumber \\
& & +\gamma_{\mu}\frac{\gamma^{\rho}k_{\rho}+m_{s}}{k^{2}-m^{2}_{s}}
\gamma_{\nu}\frac{\gamma^{\sigma}(k+p)_{\sigma}+m_{s}}{(k+p)^{2}-m^{2}_
{s}}]\}.
\end{eqnarray}
 We can see that the contributions of Pauli-Villars regulator fields to
the parity odd part (one which contains terms with three $\gamma^{'} s$)
 of $\Pi^{M}_{\mu\nu}(p)$ automatically cancel and only the contribution
 of the original spinor field remains. With the aid of the trace
formulae of 3-D $\gamma$-matrices and Feynmann parameter formulae, after
 some manupulatins and rearrangements, Eq.(5) yields

\begin{eqnarray}
\Pi^{M}_{\mu\nu}(p)&=& -2e^{2}\int^{1}_{0}dx\int\frac{d^{3}k}{(2\pi)^{3}}
\{[\frac{2k_{\mu}k_{\nu}}{(k^{2}+p^{2}x(1-x)-m^{2})^{2}}
-\frac{2k_{\mu}k_{\nu}}{(k^{2}+p^{2}x(1-x))^{2}}] \nonumber \\
& & -[\frac{g_{\mu\nu}}{k^{2}+p^{2}x(1-x)-m^{2}} -\frac{g_{\mu\nu}}{k^{2}
+p^{2}x(1-x)}]  \nonumber \\
& & +2k_{\mu}k_{\nu}\sum^{+\infty}_{n=-\infty}\frac{(-1)^{n}}{(k^{2}+p^
{2}x(1-x)-m_{n}^{2})^{2}} \nonumber \\
& & -g_{\mu\nu}\sum^{+\infty}_{n=-\infty}\frac{(-1)^{n}}{k^{2}+p^{2}
x(1-x)-m_{n}^{2}} \nonumber \\
& & -4x(1-x)(p_{\mu}p_{\nu}-p^{2}g_{\mu\nu})\sum_{n=1}^{+\infty}\frac
{(-1)^{n}}{(k^{2}+p^{2}x(1-x)-m^{2}_{n})^{2}}  \nonumber \\
& & -2x(1-x)(p_{\mu}p_{\nu}-p^{2}g_{\mu\nu})\frac{1}{(k^{2}+p^{2}x(1-x)-
m^{2})^{2}}\}  \nonumber \\
& & +2ime^{2}\varepsilon_{\mu\nu\rho}p^{\rho}\int_{0}^{1}dx\int\frac
{d^{3}k}{(2\pi)^{3}}\frac{1}{(k^{2}+p^{2}x(1-x)-m^{2})^{2}},
\end{eqnarray}
where even $n(=2r)$ and odd $n(=2s-1)$ correspod to contributions
from fermionic and bosonic regulator fields, respectively. We have also
made the replacement of momentum integration variable; $k_{\mu}
\rightarrow k_{\mu} - p_{\mu}x(1-x)$.

The integrals of first and second lines in Eq.(6) are convergent
individually and the explicit calculation shows that they cancel each
other as they must do. We note that $\int d^{3}kk_{\mu}k_{\nu}f(k^{2})=
-\frac{i}{3}g_{\mu\nu}\int d^{3}\bar{k}\bar{k}^{2}f(-\bar{k}^{2})$,
where $\bar{k}_{\mu}$ is the Euclidian three momentum and $f(k^{2})$
is any function with suitable analytic properties on the complex energy
 plane. For notational convenience we define the function $F(\bar{k}
^{2}+\bar{p}^{2}x(1-x))\equiv \sum_{n=-\infty}^{\infty} \frac{(-1)^{n}}
{\bar{k}^{2}+\bar{p}^{2}x(1-x)+m_{n}^{2}}$, where $\bar{p}^{2} \equiv
-p^{2}$. In the following let us assume the mass of regulator fields is
given by $m_{n}=Mn$, although it may not essential for our discussion.
Then the infinite series in Eq.(6) can be explicitly calculated, with
the aid of the following formulae;

\begin{eqnarray}
 f(a^{2})= \sum_{n=-\infty}^{+\infty}\frac{(-1)^{n}}{a^{2}+n^{2}} =
\frac{\pi}{a}\frac{1}{\sinh a\pi} ,\\
f^{\prime}(a^{2}) = \frac{\partial}{{\partial}a^{2}}f(a^{2}) = -\sum_
{n=-\infty}^{+\infty}\frac{(-1)^{n}}{[a^{2}+n^{2}]^{2}}, \\
 \sum_{n=1}^{+\infty}\frac{(-1)^{n}}{a^{2}+n^{2}}=-\frac{1}{2a^{2}} +
\frac{\pi}{2a}\frac{1}{\sinh a\pi}.
\end{eqnarray}
Clearly $f(a^{2})$ exponentially decreases at large $a$  and this
ensures integrals involving $f(a^{2})$ finite.

The result of the calculation is

\begin{eqnarray}
\Pi^{M}_{\mu\nu}(p)&=&-i\frac{e^{2}}{\pi^{2}}g_{\mu\nu}\int^{1}_{0}dx
\int^{\infty}_{0}d\bar{k}[\frac{1}{3}\bar{k}^{3}\frac{dF(\bar{k}^{2}+
\bar{p}^{2}x(1-x))}{d\bar{k}}+\bar{k}^{2}F(\bar{k}^{2}+\bar{p}^{2}
x(1-x))] \nonumber \\
& & +i\frac{e^{2}}{\pi^{2}M}(p_{\mu}p_{\nu}-p^{2}g_{\mu\nu})\int_{0}
^{1}dx x(1-x)\int_{0}^{\infty}d\bar{k}\{-\frac{1}{\bar{k}^{2}+\frac
{\bar{p}^{2}x(1-x)}{M^{2}}} \nonumber \\
& & +\frac{\pi}{\bar{k}^{2}+\frac{\bar{p}^{2}x(1-x)}{M^{2}}}
\frac{1}{\sinh{\pi\sqrt{\bar{k}^{2}+\frac{\bar{p}^{2}x(1-x)}{M^{2}}}}}\}
\nonumber \\
& & +i(p_{\mu}p_{\nu}-p^{2}g_{\mu\nu})\frac{e^{2}}{8{\pi}\bar{p}^{2}}
[2m+\frac{\bar{p}^{2}-4m^{2}}{\bar{p}}\arcsin{\sqrt{\frac{\bar{p}^{2}}
{\bar{p}^{2}+4m^{2}}}}] \nonumber \\
& & +i^{2}\epsilon_{\mu\nu\rho}p^{\rho}\frac{e^{2}m}{2{\pi}\bar{p}}
\arcsin{\sqrt{\frac{\bar{p}^{2}}{\bar{p}^{2}+4m^{2}}}},
\end{eqnarray}
The first integral in Eq.(10) is identically zero by partial integration.
 Therefore the gauge invariance for the regularized photon vacuum
polarization is indeed maintained independently on the choice of
regulator masses $m_{n}$. The second integral vanishes as the regulator
mass $M$ goes to infinity, since this integral is convergent both in
ultraviolet and infrared regions even when $M\rightarrow \infty$.
Finally we obtain the following result for the photon vacuum
polarization tensor $\Pi_{\mu\nu}(p)\equiv\lim_{M\rightarrow \infty}
\Pi^{M}_{\mu\nu}(p)$.

\begin{eqnarray}
\Pi_{\mu\nu}(p)&=& i(p_{\mu}p_{\nu}-p^{2}g_{\mu\nu})\frac{e^{2}}{8{\pi}
p^{2}}[2m+\frac{\bar{p}^{2}-4m^{2}}{\bar{p}}\arcsin{\sqrt{\frac{\bar{p}
^{2}}{\bar{p}^{2}+4m^{2}}}}] \nonumber \\
& &+i\epsilon_{\mu\nu\rho}p^{\rho}\frac{e^{2}m}{2{\pi}\bar{p}}\arcsin{
\sqrt{\frac{\bar{p}^{2}}{\bar{p}^{2}+4m^{2}}}}.
\end{eqnarray}
The asymptotic behaviours of $\Pi_{\mu\nu}(p)$ are, as $p\rightarrow
0$ provided $m\neq 0$

\begin{equation}
\Pi_{\mu\nu}(p) \rightarrow i(p_{\mu}p_{\nu}-p^{2}g_{\mu\nu})
\frac{e^{2}}{12\pi m}
+ i^{2}\epsilon_{\mu\nu\rho}p^{\rho}\frac{e^{2}}{4\pi},
\end{equation}
and as $m\rightarrow 0$ provided $p\neq 0$

\begin{equation}
\Pi_{\mu\nu}(p) \rightarrow i(p_{\mu}p_{\nu}-p^{2}g_{\mu\nu})\frac{e^
{2}}{16\pi \bar{p}} + i^{2}\epsilon_{\mu\nu\rho}p^{\rho}\frac{e^{2}m}
{4\bar{p}}.
\end{equation}
Now it is shown that a parity invariant Pauli-Villars regularization
metdod indeed exists and gives the same results as those obtained by the
analytic and dimensional regularization methods. Therefore at least for
the perturbative induction of Chern-Simons term in the 3-D quantum
electrodynamics three regularization methods lead to the same result.

Finally let us discuss briefly on some remaining problems. Firstly
higher-loop correction should be calculated. Secondly the extension to
the non-Abelian gauge theories where the new divergent integrals from
gauge field loops appear and cause new problems. In Refs.\cite{ALR} and
 \cite{M} it was discussed that the gauge invariant regularizations
induce the Chern-Simons term (i.e. shift the coefficient of the
Chern-Simons term). However the regularization methods in Refs.\cite{ALR}
 and \cite{M} break the parity invariance explicitly.
Therefore it is an interesting problem to investigate whether any
regularization method invariant under both gauge and parity
transformations exists for the non-Abelian gauge theories. It seems
sufficient to introduce a infinite number of parity doublets of
 massive vector fields as Pauli-Villars regulators. If this is indeed the case
it is expected that no Chern-Simons term is induced even in non-Abelian
gauge theories. These problems will be discussed in a subsequent
paper.

\vspace{5mm}
The present authour would like to thank Toru Ebihara for informing him
the work of Frolov and Slavnov.


\newpage
\begin{thebibliography}{999}

\bibitem{JT}

R. Jackiw and S. Templeton, Phys. Rev. D23, (1981) 2291;

J. F. Schonberg, Nucl. Phys. B185, (1981) 157.

\bibitem{LFC}

R. B. Laughlin, Science 242, (1988) 525;

A. L. Fettes, C. B. Hanna and R. B. Laughlin,
Phys. Rev. B39, (1989) 9679;

Y. H. Chen, B. I. Halperin, F. Wilczek and E. Witten,
Int. J. Mod. Phys B3, (1989) 1001;

N. Dorey and N. E.  Mavromatos, Nucl. Phys. B386, (1992) 614;

D. S\'{e}n\'{e}chal, Phys. Letters B297, (1992) 138.

\bibitem{WP}

E. Witten, Commun. Math. Phys. 121, (1989) 351;

A. P. Polychronakos, Ann. Phys. 203, (1990) 231.

\bibitem{DJT}

S..Deser, R. Jackiw and S. Templeton, Ann. Phys. (NY) 140, (1982) 372;

A. N. Redlich, Phys. Rev. D29, (1984) 2366;

Y. -C. Kao and M. Suzuki, Phys. Rev. D31, (1985) 2137.

\bibitem{ALR}

L. Alvarez-Gaum\'{e}, J. M. F. Labastida and A. V. Ramallo,
Nucl. Phys. B334, (1990) 103.

\bibitem{PST}

B. M. Pimentel, A. T. Suzuki and J. L. Tomazelli,
Intern. J. Mod. Phys. A7, (1992) 5307.

\bibitem{M}

C. P. Martin, Phys. Letters B241, (1990) 513.

\bibitem{DW}

R. Delbourgo and A. B. Waites, Phys. Letters B300, (1993) 241.

\bibitem{FS}

S. A. Frolov and A. A. Slavnov, Phys. Letters B309, (1993) 344;
Nucl. Phys. B411, (1994) 647.

\bibitem{NN}

R. Narayanan and H. Neugeberger, Phys. Letters B302, (1993) 62;

S. Aoki and Y. Kikukawa, Mod. Phys. Letters A8, (1993) 3517.

\end{thebibliography}
\end{document}